\documentclass{article}
\usepackage{spconf,amsmath,graphicx}
\usepackage{multirow}
\usepackage{amsfonts}
\usepackage{pifont}
\usepackage{float}
\usepackage{amsmath}
\usepackage{booktabs}

\title{SMRU: Split-and-Merge Recurrent-based UNet for Acoustic Echo Cancellation and Noise Suppression}
\name{Zhihang Sun\textsuperscript{1,2}, Andong Li\textsuperscript{1}, Rilin Chen\textsuperscript{1}, Hao Zhang\textsuperscript{1}, Meng Yu\textsuperscript{1}, Yi Zhou\textsuperscript{2}, Dong Yu\textsuperscript{1}}
\address{\textsuperscript{1}Tencent AI Lab \\
\textsuperscript{2}School of Communications and Information Engineering, \\ Chongqing University of Posts and Telecommunications, Chongqing, China
}
\begin{document}
%
\maketitle
%
\begin{abstract}
The proliferation of deep neural networks has spawned the rapid development of acoustic echo cancellation and noise suppression, and plenty of prior arts have been proposed, which yield promising performance. Nevertheless, they rarely consider the deployment generality in different processing scenarios, such as edge devices, and cloud processing. To this end, this paper proposes a general model, termed SMRU, to cover different application scenarios. The novelty lies in two-fold. First, a multi-scale band split layer and band merge layer are proposed to effectively fuse local frequency bands for lower complexity modeling. Besides, by simulating the multi-resolution feature modeling characteristic of the classical UNet structure, a novel recurrent-dominated UNet is devised. It consists of multiple variable frame rate blocks, each of which involves the causal time down-/up-sampling layer with varying compression ratios and the dual-path structure for inter- and intra-band modeling. The model is configured from 50 M/s to 6.8 G/s in terms of MACs, and the experimental results show that the proposed approach yields competitive or even better performance over existing baselines, and has the full potential to adapt to more general scenarios with varying complexity requirements.
\end{abstract}
\section{Introduction}
\label{sec:intro}

Annoying acoustic echo and environmental noise are ubiquitous in real-time communication (RTC) systems, leading to hurdles in intelligibility and overall low audio quality. Digital Signal Processing (DSP)-based methods for linear acoustic echo cancellation (AEC) were widely adopted in RTC scenarios~{\cite{soo1990multidelay, enzner2006frequency}}. Nonetheless, these classical approaches cannot effectively cancel acoustic echo and struggle to maintain high speech quality in relatively low signal-to-noise ratios (SNRs) and double-talk scenarios. Recent years have witnessed
the proliferation of deep neural networks (DNNs) and dozens of DNN-based AEC algorithms have been proposed, which can be roughly categorized into two classes, namely hybrid~{\cite{zhao2022deep,franzen2022deep,zhang2022multi,sun2022explore}} and fully neural network-based systems~{\cite{zhang2019deep,westhausen2021acoustic, zheng2023real}}, depending on whether the linear-AEC is involved as a prior for later DNN processing.

Regarding the neural network topology in the AEC task, an intuitive tactic is to transfer the network structures from other front-end tasks like speech enhancement~{\cite{tan2019learning,kim2021se,yu2022dual,dang2022dpt}}. For example, in~{\cite{zhang2022multi}}, a classical UNet-style structure was utilized, in which convolution-based encoder and decoder are adopted for feature extraction and target spectrum recovery, and stacked LSTM layers serve as the bottleneck for temporal and frequency modeling. In~{\cite{sun2022explore}}, a dual-path transformer structure was devised to effectively grasp global relations. Despite the promising performance these works have achieved, the computational complexity is usually prohibitive and they might be quite difficult to deploy in edge devices. Besides, some operators like self-attention may require large time buffers, which can bring laborious optimization costs. Considering that, an important question arises: \textit{how to devise an AEC network that encompasses different complexity and is also felicitous to adapt to real-time scenarios.}
\begin{figure}[t]
  \centering
\includegraphics[width=0.9\linewidth]{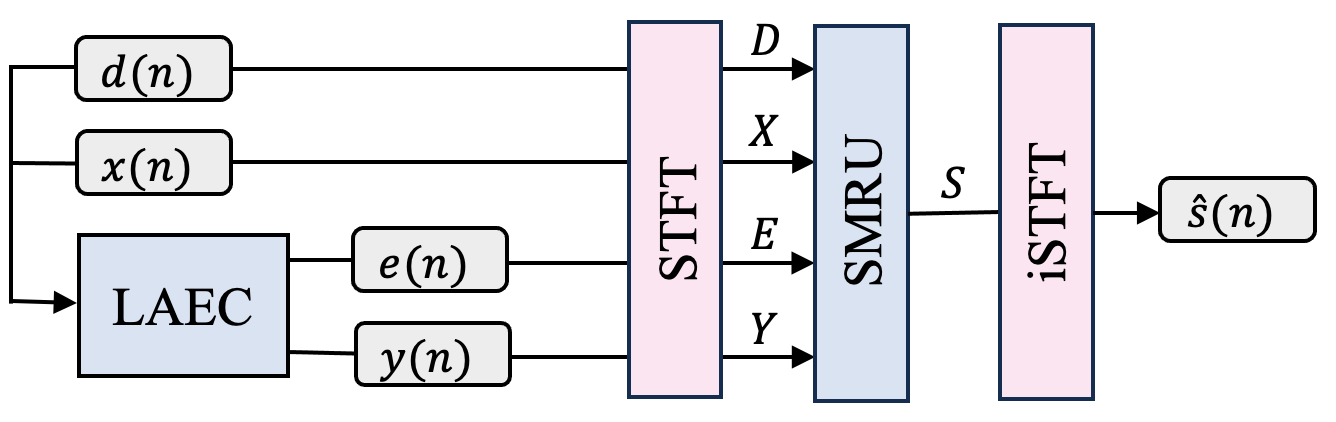}
  \caption{Overview diagram of the proposed hybrid AEC system.}
 \label{fig:aec-framework}
 \vspace{-0.05cm}
\end{figure}

\begin{figure*}[t]
  \centering
\includegraphics[width=1.0\linewidth]{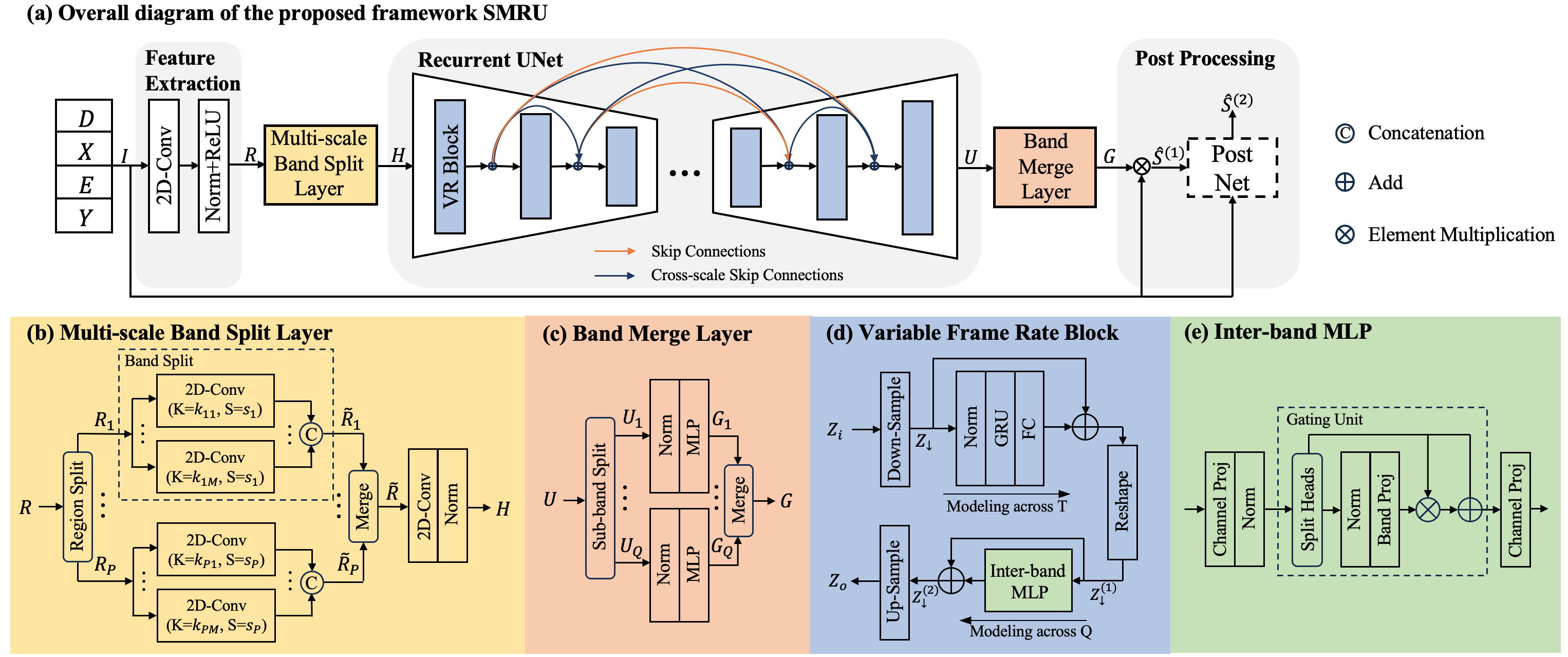}
  \caption{Architecture of the proposed SMRU. Different modules are indicated with different colors for better illustrations. (a) Overall diagram of the proposed SMRU. (b) Detail structure of the multi-scale band split layer. (c) Detail structure of the band merge layer. (d) Detail structure of the variable frame rate block. (e) Detail structure of the inter-band MLP.}
  \vspace{-0.0cm}
  \label{fig:smru}
\end{figure*}

Note that most of the previous works require the number of frames to be unaltered in the forward process to follow the causality principle, and the time down-sampling/up-sampling (DS/US) operations are often not allowed, which prunes to result in high computational complexity. More recently, a causal-guaranteed DS/US strategy was proposed by future frame prediction~{\cite{hao2022fast,chen2023ultra}}, leading to notably decreased complexity and mild performance degradation. Besides, in~{\cite{yu2022high}}, a band-split strategy was proposed to manually merge neighboring frequency sub-bands and can effectively decrease the cost aroused by modeling in a large frequency dimension. Therefore, we believe that it should be significant to flexibly modulate the time and frequency dimensions and meanwhile sustain the causality characteristic for a real-time AEC framework.

In this regard, we propose \underline{\textbf{S}}plit-and-\underline{\textbf{M}}erge \underline{\textbf{R}}ecurrent-based \underline{\textbf{U}}Net dubbed \textbf{SMRU}, the first UNet-style recurrent-dominated framework for AEC and noise suppression to our best knowledge. The whole framework adopts the UNet topology structure, in which the encoder part follows the from-fine-to-coarse principle and vice versa for the decoder, and the skip connections are utilized for feature recalibration. Different from preliminary convolution-based UNet works, here the ``coarse/fine" lies in different temporal resolution instead of frequency size, \textit{i.e.}, multi-level causal time DS/US operations are utilized in the encoder and decoder, respectively, enabling multi-scale temporal modeling and also notably reducing the computation complexity. Within each module, a dual-path structure was devised, where the RNN excavates the temporal relations and an MLP-based band shuffler is adopted for global band modeling~{\cite{tu2022maxim}}. Benefiting from the multi-scale time compression method, the proposed model enjoys better flexibility in computational complexity control. In this paper, the proposed model can cover from 50 M/s to 6.8 G/s in terms of MACs, and both quantitative and qualitative results manifest the performance superiority of the proposed method.

The rest of the paper is organized as follows. Sec.~{\ref{sec:proposed-system}} presents the proposed framework. Sec.~{\ref{sec:experimental-setup}} and Sec.~{\ref{sec:experimental-results}} namely give the experimental setups and results. Some conclusions are drawn in Sec.~{\ref{sec:conclusion}}.

\section{Proposed Method}
\label{sec:proposed-system}

The overview diagram of the proposed hybrid AEC system is shown in Figure~{\ref{fig:aec-framework}}. The input consists of the received microphone mixture signal \(d(n)\), the reference far-end signal \(x(n)\), the error signal \(e(n)\), and the linear echo \(y(n)\) generated by LAEC algorithm~{\cite{kuech2014state}}, respectively, and $n$ denotes the time sample index. All signals are then converted to the time-frequency (T-F) domain and fed into the SMRU. The framework of the proposed SMRU is shown in Figure~{\ref{fig:smru}}. The real and imaginary parts of the four input spectra are concatenated along the channel axis to yield the input feature $I\in\mathbb{R}^{8\times T\times F}$, where \(T\) and \(F\) denote the number of the frames and frequency bins, respectively. The input first passes a 2D convolution layer to generate an initial feature map $R\in\mathbb{R}^{E\times T\times F}$, where $E$ denotes the embedding dimension. Similar to~{\cite{yu2022high}}, the feature map is split into sub-bands by the band split layer (see Figure~{\ref{fig:smru}}(b)) to compress the frequency dimension. The compressed feature is then fed into the proposed recurrent UNet for multi-scale modeling. After that, the band merge layer (see Figure~{\ref{fig:smru}}(c)) is adopted for filter estimation. A lightweight postnet is optional and can be utilized for further post-processing. We will illustrate each module in detail below.

\subsection{Split and Merge}

To alleviate the computational burden caused by a large frequency dimension, we manually divide all frequency bins into three frequency regions, representing low, mid, and high regions. For each region, different multi-scale convolution sets are used to split and compress the frequency bins to a unified embedding dimension $E$. After the UNet modeling, each sub-band is converted to its original size and merged. The split and merge pattern allows the feature stream to maintain a relatively low dimension, while multi-scale convolution sets can introduce richer inter-band information. 

\subsubsection{Multi-scale band split layer}

Figure~{\ref{fig:smru}}(b) shows the detail of the band split layer. The input feature is split along the frequency dimension into $P$ regions. Each region feature $R_{p}$ is processed by a set of 2D convolutions with different kernel sizes and the same output channels. They are subsequently concatenated to obtain a compressed 3D representation $\tilde{R}_{p}$, where subscript $p\in\left\{1,\cdots,P\right\}$. All $\tilde{R}_{p}$ are merged into $\tilde{R} \in \mathbb{R}^{(M \times E) \times T \times Q}$, where $M$ denotes the convolution scales, and $Q$ denotes the number of sub-bands after compression. A 2D convolution is then applied to reduce the embedding dimension of $\tilde{R}$ from $M\times E$ to $E$. Collectively, the process can be formulated as:
\begin{small}
\vspace{-0.0cm}
\begin{equation}
\left\{{R}_{1},\cdots,{R}_{P}\right\} = Region\text{-}split(R),
\end{equation}
\end{small}
\vspace{-0.0cm}
\begin{small}
\begin{equation}
{\tilde{R}}_{p} = Cat(Conv_{K=k_{p1},S=s_{p}}(R_{p}),\cdots,Conv_{K=k_{pM},S=s_{p}}(R_{p})),
\vspace{-0.0cm}
\end{equation}
\end{small}
\begin{equation}
{H}=Norm(Conv(Merge(\tilde{R}_{1},\cdots,\tilde{R}_{P}))),
\vspace{-0.00cm}
\end{equation}
where $Cat\left(\cdot\right)$ and $Merge\left(\cdot\right)$ refer to the concatenation operation along the channel and frequency axes, respectively. $H\in\mathbb{R}^{E\times T\times Q}$ denotes the output from the split layer. Please note that different regions adopt different strides in their convolution sets, as the frequency compression ratio varies for each $R_p$. Recall that in~{\cite{yu2022high}}, the band split/merge operations are implemented with \textit{for-loop}, and we notice a higher implementation efficiency with convolution operations thanks to the internal optimization of Pytorch platform. When we adopt the convolutions in the split layer but not the merge part, only neglectable performance degradations are observed in our internal trials.

\subsubsection{Band merge layer}
Figure~{\ref{fig:smru}(c) shows the internal structure of the band merge layer. To be specific, the output from the recurrent UNet is termed as $U$. It is split into $Q$ sub-band features, and each sub-band feature is fed into a normalization layer and a separate multilayer perceptron (MLP) layer to estimate complex-valued T-F masks ${G}_{q}$, where subscript $q\in\left\{1,\cdots, Q \right\}$. Finally, all ${G}_{q}$ are merged along the frequency axis to obtain the estimated T-F mask $G\in \mathbb{R}^{8\times T \times F}$, which is combined with $I$ for target spectrum filtering. The process can be given by:
\vspace{-0.0cm}
\begin{equation}
\left\{U_{1},\cdots, U_{Q}\right\} = Sub\text{-}band\text{-}Split\left(U\right),
\vspace{-0.0cm}
\end{equation}
\begin{equation}
{G}_{q} = MLP_{q}\left(Norm\left(U_{q}\right)\right),
\vspace{-0.0cm}
\end{equation}
\begin{equation}
{G} = Merge\left({G}_{1},\cdots,{G}_{Q}\right),
\vspace{-0.0cm}
\end{equation}
\begin{equation}
\hat{S}^{\left(1\right)} = \sum_{i}^{8}I_{i}\otimes G_{i},
\vspace{-0.0cm}
\end{equation}
where $\otimes$ denotes the element multiplication\ operator, and $\hat{S}^{\left(1\right)}$ denotes the target estimation after filtering.

\subsection{Recurrent UNet}
\subsubsection{Variable frame rate block}
The proposed recurrent UNet comprises multiple basic blocks with variable frame rates (VR) due to different time DS/US operations. Taking the encoder process as an example, the internal structure of each VR block is shown in Figure~{\ref{fig:smru}}(d). We may as well denote the input feature as $Z_{i}$. It is first passed to a causal time DS operation to obtain a squeezed version with a lower frame rate feature stream termed as $Z_{\downarrow}$. Then a dual-path module is utilized to model the intra-band and inter-band relations, respectively. Finally, the causal time US layer is adopted to recover the feature back to its original frame rate, termed as $Z_{o}$. The above-mentioned process can be summarized as:
\vspace{-0.1cm}
\begin{equation}
{Z}_{\downarrow}=TimeDownSample(Z_{i}),
\vspace{-0.0cm}
\end{equation}
\begin{equation}
{Z}_{\downarrow}^{(1)} = Reshape(FC(GRU(Norm(Z_{\downarrow})))+Z_{\downarrow}),
\vspace{-0.0cm}
\end{equation}
\begin{equation}
{Z}_{\downarrow}^{(2)} = Inter\text{-}band\text{-}MLP({Z}_{\downarrow}^{(1)})+{Z}_{\downarrow}^{(1)},
\vspace{-0.0cm}
\end{equation}
\begin{equation}
Z_{o}=TimeUpSample({Z}_{\downarrow}^{(2)}),
\vspace{-0.0cm}
\end{equation}
where ${Z}_{\downarrow}^{(1)}$ and ${Z}_{\downarrow}^{(2)}$ denote the outputs from inter-band and intra-band modeling, respectively. $Reshape(\cdot)$ denotes the transpose operation.

\subsubsection{Causal time Down-Sample and Up-Sample layer}
The Down-Sample layer is implemented using non-overlapped 1D causal convolution and $Reshape(\cdot)$ operation. In concrete, we merge the embedding and sub-band dimensions and pass a causal 1D convolution layer to obtain a down-sampled version, \textit{i.e.}, $\mathbb{R}^{\left(E\times Q\right)\times T}\mapsto\mathbb{R}^{\left(E\times Q\right)\times \left(T/\lambda\right)}$, where $\lambda$ denotes the time compression ratio, and the kernel size and stride are set to the same value to the compression ratio to keep causality. In the Up-Sample layer, the input is first interpolated in the time axis, and a 1D point-wise convolution layer is then used. The causality is guaranteed by future frame prediction. Due to the space limit, we refer the readers to~{\cite{chen2023ultra}} for more details.

\subsubsection{Inter-band MLP shuffler}
The Inter-band MLP shuffled is inspired by the gMLP proposed in {\cite{liu2021pay}}. It consists of channel and band projections and uses split head and multiplicative gating for global inter-band modeling. The channel dimension of the feature map is doubled using 1D convolution in the first channel projections. Band attention is achieved through the Gating Unit~{\cite{tu2022maxim}}, which evenly divides the feature map into two parts along the channel dimension. One part is modeled inter-band along the time axis in the band projections using 1D convolution, and then multiplicative gate with the other part. Finally, in the last channel projections, we reshape the feature map and apply 1D convolution again.

\subsubsection{Cross-scale skip connections}
In addition to the standard skip connections in UNet, we also introduce cross-scale skip connections, as shown by the blue curve in Figure~{\ref{fig:smru}}. We adopt a connection mode similar to dense connections {\cite{huang2017densely}}, but instead of channel-wise concatenation, we sum them after normalization. By doing so, nearly no extra computational overhead is introduced. We observe that the strategy can effectively improve performance, which will be revealed in Sec.~{\ref{sec:ablation-study}}.

\subsection{Post-processing module}
Despite the effectiveness of the proposed model, the residual noise components may still exist. To further suppress the remaining noise, a lightweight postnet is often cascaded for post-processing. Similar to~{\cite{schroter2022deepfilternet}}, it consists of several GRU layers and a group linear layer to estimate deep filter coefficients for deep filtering. The complexity of the adopted postnet is only 30 M/s in terms of MACs, which can be overall negligible.

\subsection{Loss function}
We adopt the Mean Absolute Error (MAE) loss, which is formulated as 
\vspace{-0.0cm}
\begin{equation}
\label{eqn:eq2}
\mathcal{L}_{MAE} = MAE(\hat{S}_{R}, S_{R})+MAE(\hat{S}_{I}, S_{I})+MAE(|\hat{S}|, |S|),
\end{equation}
where $\left\{\hat{S}_{R}, \hat{S}_{I}, \hat{\left|S\right|}\right\}$ refer to the real, imaginary, and magnitude parts of the estimated spectrum, respectively, and $\left\{{S}_{R}, {S}_{I}, \left|{S}\right|\right\}$ refer to that of the target version. To effectively suppress the echo, the echo-aware loss $\mathcal{L}_{echo}$~{\cite{zhang2022multi}} is also adopted. Besides, when the near-end speech is absent, we expect the model to suppress the output as much as possible. To this end, the VAD-oriented loss $\mathcal{L}_{vad}$ is proposed, given by 
\vspace{-0.0cm}
\begin{equation}
\label{eqn:eq1}
\mathcal{L}_{vad} = 10\log_{10}(||\hat{S} \times (1-{\mathbb{I}}_{vad})||_{2}^{2}+\epsilon),
\vspace{-0.0cm}
\end{equation}
where \(\hat{S}\) represents the predicted spectrum, \({\mathbb{I}}_{vad}\) is the VAD label of the near-end speech, and \(\epsilon\) is used to prevent over-suppression of the target speech, which we empirically set to 0.1. The final loss is thus given by
\vspace{-0.0cm}
\begin{equation}
\label{eqn:eq1}
\mathcal{L} = \mathcal{L}_{MAE}+0.1\mathcal{L}_{echo}+\beta\mathcal{L}_{vad},
\vspace{-0.0cm}
\end{equation}
where $\beta$ is set to balance the capability between echo suppression and near-end speech preservation, whose impact will be shown in Sec.~{\ref{sec:ablation-study}}.

\section{Experimental setup}
\label{sec:experimental-setup}

\subsection{Data preparation}
In the training dataset, clean clips are randomly sampled from the \textit{train-clean-100} and \textit{train-clean-360} subsets of Librispeech~{\cite{panayotov2015librispeech}}. Environmental noises are sampled from the DNS-Challenge~{\cite{reddy2021interspeech}}. The echo data are simulated by convolving clean speech with room impulse responses (RIRs) from the SLR28 dataset {\cite{ko2017study}}. The simulated SNR and signal-to-echo ratio (SER) are sampled from -5 to 15 dB. The scenario proportions of far-end single talk (ST-FE), near-end single talk (ST-NE), and double talk (DT) are set to 10\%, 25\%, and 65\%, respectively. Besides, noise can be absent in 10\% of the data. The duration of the training set is around 530 hours, and 5\% of the data in the training set are picked out for model validation.

The test set is simulated by the same method with different data sources. Clean audios are sourced from the \textit{test-clean} subset of Librispeech. Noise audios are selected from the same dataset as the training set, with no data overlap. The echo data are obtained from real recorded echoes from the AEC Challenge~{\cite{cutler2022icassp}}. SERs and SNRs of $-5$ dB, $5$ dB, $15$ dB, $+\infty$ and $-\infty$ are included in the test set. A SER of $+\infty$ corresponds to the ST-NE scenario, while a SER of $-\infty$ corresponds to the ST-FE scenario. The total duration of the test set was around 10 hours. In addition, we use the blind test set of the AEC Challenge~{\cite{cutler2022icassp}} to investigate the generalization capability of models.

\subsection{Implementation details}
A state-space-based linear filter is used to estimate the error signal $e(n)$ and linear echo $y(n)$ in the LAEC~{\cite{kuech2014state}}. The window length for STFT and iSTFT is set to 20 ms, with an overlap of 10 ms. The number of VR blocks is 12, with 6 blocks set in the encoder and decoder, respectively. The time compression ratios $\lambda$ are set as $\left\{1, 2, 4, 8, 16, 32, 32, 16, 8, 4, 2, 1\right\}$. In our experiments, we can adjust the model complexity by changing the embedding dimension $E$. For $E=10$ and $E=200$, the computational complexity of the model (without post-processing module) can be 50 M/s and 6.8 G/s, respectively, which are adequate to cover both resource-limited and offline scenarios. For the multi-scale band split layer, the number of regions $P$ is set to 3, with each region corresponding to 20, 60, and 81 frequency bins, respectively. The stride of the 2D convolution sets for the 3 regions is set to $\left\{(2, 4), (2, 10), (2, 20)\right\}$, and the corresponding kernels $K_{1}$, $K_{2}$, $K_{3}$ are respectively set to $\left\{(1, 4), (1, 8), (1, 12)\right\}$, $\left\{(1, 10), (1, 20), (1, 30)\right\}$, $\left\{(1, 20), (1, 30), (1, 40)\right\}$. The Adam optimizer is adopted with an initialized learning rate of 0.001 and a decay coefficient of 0.99. Each model is trained for 200 epochs with a batch size of 16 in the utterance level.

\subsection{Evaluation metrics}
For the synthetic test set, the ST-NE and DT scenarios are evaluated using scale-invariant SNR (SI-SNR) {\cite{luo2019conv}} and wide-band perceptual evaluation of speech quality (WB-PESQ) {\cite{rec2005p}}, while the ST-FE scenario is evaluated using echo return loss enhancement (ERLE) {\cite{theodoridis2013academic}}. For the blind test set, we use the AECMOS metric {\cite{purin2022aecmos}}.

\begin{table}[t]
    \caption{The objective results of proposed SMRU and different baselines in the test set. \textbf{BOLD} indicates the best score.}
    \vspace{5pt}
    \normalsize
    \setlength{\tabcolsep}{3pt}
    \centering
    \resizebox{1.0\columnwidth}{!}{
        \begin{tabular}{l|cc|cc|cc|c}
            \toprule
            \multirow{2}*{Models} &MACs &\multirow{2}*{RTF} 
            &\multicolumn{2}{c|}{DT}
            &\multicolumn{2}{c|}{ST-NE}
            &\multirow{1}*{ST-FE}\\
            &(G/s) & &SI-SNR &PESQ &SI-SNR &PESQ &ERLE \\
            \midrule
            NSNet &0.13 &\textbf{0.0114} &10.01 &1.82 &12.94 &2.08 &52.16 \\
            DeepFilterNet &0.24 &0.0347 &11.48 &2.11 &14.01 &2.31 &55.94 \\
            DTLN &0.46 &0.0351 &11.65 &2.08 &13.02 &2.14 &\textbf{67.39} \\
            BSRNN &1.38 &0.2087 &12.92 &2.34 &14.67 &2.48 &55.03 \\
            FastFullSubNet &1.75 &0.1008 &12.64 &2.25 &14.18 &2.35 &50.61 \\
            \midrule
            SMRU-T &\textbf{0.05} &0.0291 &11.17 &1.97 &12.90 &2.08 &52.93\\
            SMRU-S &0.11 &0.0354 &11.76 &2.09 &13.58 &2.21 &52.87\\
            \hspace{2em}+PostNet &0.14 &0.0496 &12.29 &2.17 &13.97 &2.29 &52.44\\
            SMRU-L &1.03 &0.0972 &13.28 &2.35 &14.77 &2.48 &57.18\\
            SMRU-H &6.83 &0.3452 &\textbf{14.11} &\textbf{2.50} &\textbf{15.65} &\textbf{2.65} &58.91\\
            \bottomrule
    \end{tabular}}
    \label{tbl:results-comparison}
    \vspace{-0.00cm}
\end{table}

\begin{figure}[t]
  \centering
  \includegraphics[width=1.0\linewidth]{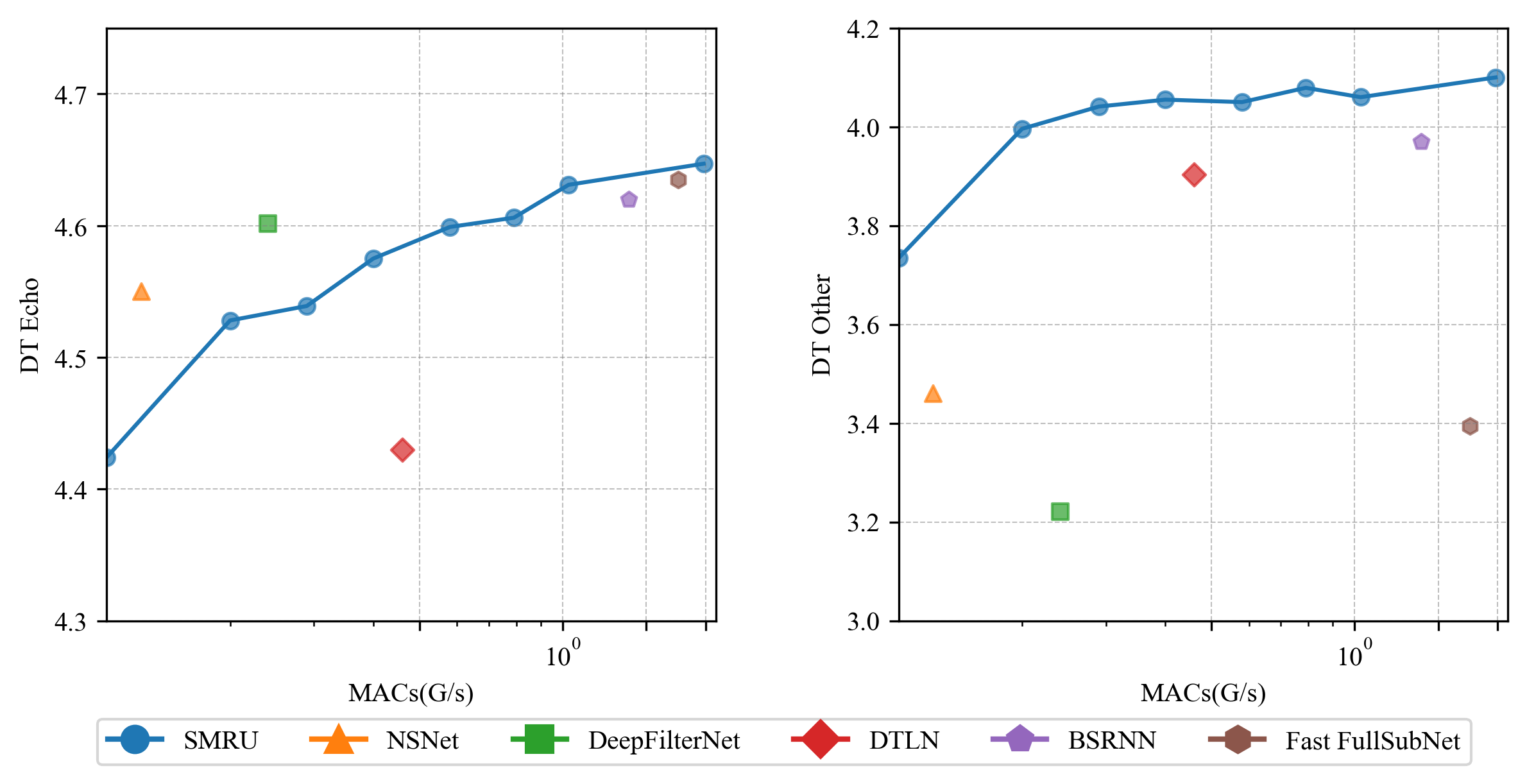}
  \caption{AECMOS metrics of the blind test set  under the DT scenario.}
  \vspace{-0.0cm}
  \label{fig:aec-mos}
\end{figure}

\begin{figure}[t]
  \centering
  \includegraphics[width=1.0\linewidth]{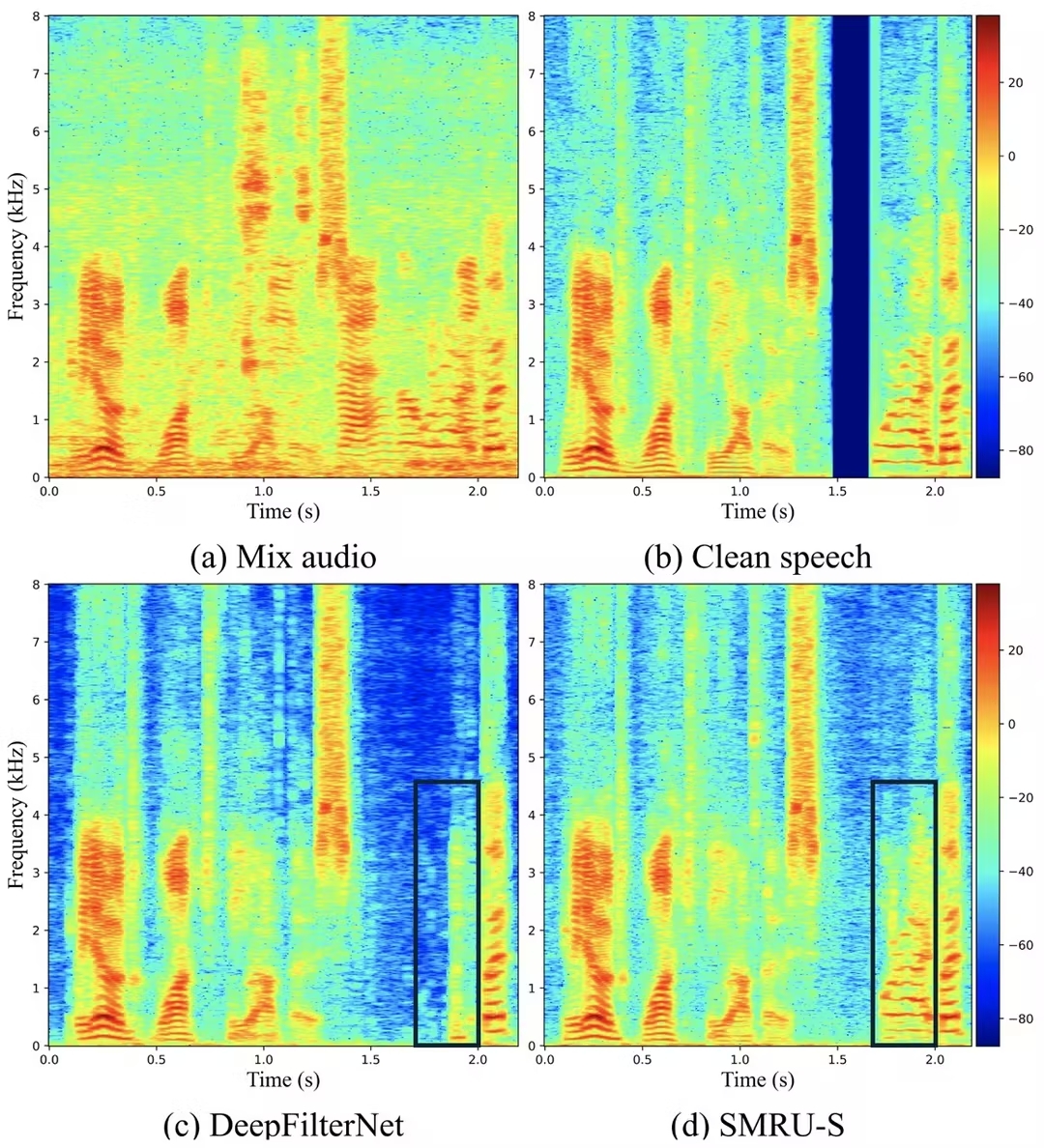}
  \caption{Spectrum visualizations of an example. (a) Mix audio. (b) Target near-end speech. (c) Estimated spectrum processed by DeepFilterNet. (d) Estimated spectrum processed by SMRU-S.}
\vspace{-0.0cm}
  \label{fig:speech-visualizations}
\end{figure}

\section{Experimental results}
\label{sec:experimental-results}

\subsection{Result comparisons with baselines}

We compare the proposed SMRU with five advanced baselines on the test set, and quantitative results are shown in Table~{\ref{tbl:results-comparison}}. Four modes of SMRU are investigated, namely tiny (T), small (S), large (L), and huge (H), with the complexity varying from 50 M/s to 6.83 G/s, in terms of MACs. Baseline methods include NSNet {\cite{cutler2022icassp}}, DTLN {\cite{westhausen2021acoustic}}, DeepFilterNet, FastFullSubNet, and BSRNN. The latter three models were originally proposed for the speech enhancement task, and we adapt them to AEC task by using the same input as the proposed method. The real-time factor (RTF) is measured on an Intel Core (TM) i7-9750H CPU clocked at 2.60 GHz. From the table, several observations can be made. First, with the increase in computational complexity, the objective metric scores of the proposed method are gradually improved, where the tiny version achieves overall better performance over NSNet but only with around one-third percentage in complexity. Moreover, although DTLN performs well in the ST-FE scenario, it lacks capability in other scenarios. In DT and ST-NE scenarios, DTLN performs worse compared to SMRU-S, which has only a quarter of its complexity. For the large version, SMRU further outperforms BSRNN and FastFullSubNet with less complexity. It fully validates the superiority of the proposed method. Besides, when a PostNet is adopted, notable improvements can be observed in both DT and ST-NE cases and just slight degradation in ERLE for the ST-FE case, which validates the effectiveness of the post-processing. Finally, compared with BSRNN and FastFullSubNet, SMRU-L enjoys a notably lower RTF, which can be attributed to the proposed UNet structure with a multi-level time sampling strategy.

Figure~{\ref{fig:aec-mos}} shows the AECMOS results on the AEC Challenge ICASSP 2022 blind test set for different approaches under the DT scenario. The MOS change trend of SMRU with different complexities is shown by the blue curve. One can see that SMRU provides the best DT performance. The MOS of DT Echo for NSNet and DeepFilterNet is slightly better than that of the same complexity SMRU, but their near-end speech preservation is poor, resulting in low MOS of DT Other. Therefore, SMRU can strike well trade-off between far-end echo suppression and near-end speech preservation.

Figure~{\ref{fig:speech-visualizations}} shows the spectrum visualization of an example case. (a)-(d) denote mix, near-end, estimation of DeepFilterNet, and SMRU-S, respectively. It can be observed that the result of the DeepFilterNet may over-suppress the voiced regions after the silent segment, while the proposed SMRU can better preserve the harmonic structure of this target speech.

\subsection{Ablation study}
\label{sec:ablation-study}
Ablation studies are conducted on SMRU-S to investigate the effects of using multi-scale band split layer and cross-scale skip connections in the test set. For the case without the multi-scale band split layer, the single-scale convolution is used for band splitting, \textit{i.e.}, the number of convolutions in the convolution set for each region is set to 1. The results in Table~{\ref{tbl:ablation}} show that removing the multi-scale band split layer can lead to significant performance degradation, as the single-scale convolution cannot provide frequency representations at different resolutions. Besides, the cross-scale skip connections can also provide a notable performance improvement by only introducing minimal computational overhead.

Table~{\ref{tbl:vad-related}} shows the performance of different $\beta$ values in the loss function. When $\beta=0.0002$, both PESQ in DT and ST-NE scenarios and ERLE in ST-FE scenario show an improvement over $\beta=0$. However, when $\beta$ further increases, the model exhibits better capability in far-end echo suppression, \textit{i.e.}, a higher ERLE score, but at the cost of more speech distortion, \textit{i.e.}, lower PESQ and SI-SNR scores. Due to space constraints, we do not traverse more $\beta$ options, and $\beta=0.0002$ seems adequate to well balance between echo cancellation and target speech preservation. 

\renewcommand\arraystretch{1.0}
\begin{table}[t]
    \caption{Ablation study on the proposed SMRU-S. Cond. 1 represents the condition of using a multi-scale band split layer, while Cond. 2 represents the condition of using cross-scale skip connections.}
    \vspace{5pt}
    \normalsize
    \setlength{\tabcolsep}{3pt}
    \centering
    \resizebox{1.0\columnwidth}{!}{
        \begin{tabular}{l|c|c|cc|cc|c}
            \toprule
            \multirow{2}*{Model} & \multirow{2}{*}{\shortstack{Cond. 1}} & \multirow{2}{*}{\shortstack{Cond. 2}}
            &\multicolumn{2}{c|}{DT}
            &\multicolumn{2}{c|}{ST-NE}
            &\multirow{1}*{ST-FE}\\
            & & &SI-SNR &PESQ &SI-SNR &PESQ &ERLE \\
            \midrule
            \multirow{3}{*}{SMRU-S} & \checkmark & \ding{55} &11.69 &2.06 &13.39 &2.17 &52.51 \\
            & \ding{55} & \checkmark &11.54 &2.02 &13.21 &2.12 &45.59 \\
            & \checkmark & \checkmark &\textbf{11.76} &\textbf{2.09} &\textbf{13.58} &\textbf{2.21} &\textbf{52.87} \\
            \bottomrule
        \end{tabular}}
        \label{tbl:ablation}
\vspace{-0pt}
\end{table}

\renewcommand\arraystretch{1.0}
\begin{table}[t]
    \caption{Ablation study on the weight \(\beta\) of the proposed VAD-oriented loss.}
    \vspace{5pt}
    \normalsize
    \setlength{\tabcolsep}{3pt}
    \centering
    \resizebox{0.9\columnwidth}{!}{
        \begin{tabular}{l|c|cc|cc|c}
            \toprule
            \multirow{2}*{Model} & \multirow{2}{*}{$\beta$}
            &\multicolumn{2}{c|}{DT}
            &\multicolumn{2}{c|}{ST-NE}
            &\multirow{1}*{ST-FE}\\
            & &SI-SNR &PESQ &SI-SNR &PESQ &ERLE \\
            \midrule
            \multirow{4}{*}{SMRU-S} & 0 &\textbf{11.85} &2.08 &13.51 &2.19 &50.73 \\
            & 0.0002 &11.76 &\textbf{2.09} &\textbf{13.58} &\textbf{2.21} &52.87 \\
            & 0.0005 &11.76 &2.06 &13.45 &2.18 &55.56 \\
            & 0.001 &11.60 &2.04 &13.21 &2.16 &\textbf{62.81} \\
            \bottomrule
        \end{tabular}}
        \label{tbl:vad-related}
\end{table}

\section{Conclusion}
\label{sec:conclusion}
In this paper, we propose SMRU, a UNet-based fundamental model for echo cancellation and noise suppression task. To enable more flexible computational complexity control, we explore modulating both frequency and time dimensions. For the former, the multi-scale band split layer and band merge layer are introduced to effectively decrease the modeling complexity in the frequency domain. For the latter, we introduce the variable frame rate block as the basic unit to model both intra-/inter-band, while also effectively decreasing the computational complexity via different causal time down-/up-sampling rates. With these tactics together, we control the overall computational complexity from 50 M/s to 6.8 G/s in MACs, which is adequate to cover both resource-limited and cloud-processing scenarios. Both quantitative and qualitative results reveal the superiority of the proposed approach over existing advanced baselines. In future work, we plan to extend the proposed SMRU to more related tasks, \emph{e.g.}, dereverberation and multi-channel speech enhancement.

\bibliographystyle{IEEEbib}
\bibliography{main}

\begin{thebibliography}{10}

\bibitem{soo1990multidelay}
J-S Soo and Khee~K Pang,
\newblock ``Multidelay block frequency domain adaptive filter,''
\newblock {\em IEEE Trans. Acoust. Speech Signal Process.}, vol. 38, no. 2, pp. 373--376, 1990.

\bibitem{enzner2006frequency}
Gerald Enzner and Peter Vary,
\newblock ``Frequency-domain adaptive kalman filter for acoustic echo control in hands-free telephones,''
\newblock {\em Signal Process.}, vol. 86, no. 6, pp. 1140--1156, 2006.

\bibitem{zhao2022deep}
Haoran Zhao, Nan Li, Runqiang Han, Lianwu Chen, Xiguang Zheng, Chen Zhang, Liang Guo, and Bing Yu,
\newblock ``A deep hierarchical fusion network for fullband acoustic echo cancellation,''
\newblock in {\em Proc. IEEE Int. Conf. Acoust. Speech Signal Process.}, 2022, pp. 9112--9116.

\bibitem{franzen2022deep}
Jan Franzen and Tim Fingscheidt,
\newblock ``Deep residual echo suppression and noise reduction: A multi-input fcrn approach in a hybrid speech enhancement system,''
\newblock in {\em Proc. IEEE Int. Conf. Acoust. Speech Signal Process.}, 2022, pp. 666--670.

\bibitem{zhang2022multi}
Shimin Zhang, Ziteng Wang, Jiayao Sun, Yihui Fu, Biao Tian, Qiang Fu, and Lei Xie,
\newblock ``Multi-task deep residual echo suppression with echo-aware loss,''
\newblock in {\em Proc. IEEE Int. Conf. Acoust. Speech Signal Process.}, 2022, pp. 9127--9131.

\bibitem{sun2022explore}
Xingwei Sun, Chenbin Cao, Qinglong Li, Linzhang Wang, and Fei Xiang,
\newblock ``Explore relative and context information with transformer for joint acoustic echo cancellation and speech enhancement,''
\newblock in {\em Proc. IEEE Int. Conf. Acoust. Speech Signal Process.}, 2022, pp. 9117--9121.

\bibitem{zhang2019deep}
Hao Zhang, Ke~Tan, and DeLiang Wang,
\newblock ``Deep learning for joint acoustic echo and noise cancellation with nonlinear distortions.,''
\newblock in {\em Proc. Interspeech}, 2019, pp. 4255--4259.

\bibitem{westhausen2021acoustic}
Nils~L Westhausen and Bernd~T Meyer,
\newblock ``Acoustic echo cancellation with the dual-signal transformation lstm network,''
\newblock in {\em Proc. IEEE Int. Conf. Acoust. Speech Signal Process.}, 2021, pp. 7138--7142.

\bibitem{zheng2023real}
Chengyu Zheng, Yuan Zhou, Xiulian Peng, Yuan Zhang, and Yan Lu,
\newblock ``Real-time speech enhancement with dynamic attention span,''
\newblock in {\em Proc. IEEE Int. Conf. Acoust. Speech Signal Process.}, 2023, pp. 1--5.

\bibitem{tan2019learning}
Ke~Tan and DeLiang Wang,
\newblock ``Learning complex spectral mapping with gated convolutional recurrent networks for monaural speech enhancement,''
\newblock {\em IEEE Trans. Audio Speech Lang. Process.}, vol. 28, pp. 380--390, 2019.

\bibitem{kim2021se}
Eesung Kim and Hyeji Seo,
\newblock ``Se-conformer: Time-domain speech enhancement using conformer.,''
\newblock in {\em Proc. Interspeech}, 2021, pp. 2736--2740.

\bibitem{yu2022dual}
Guochen Yu, Andong Li, Chengshi Zheng, Yinuo Guo, Yutian Wang, and Hui Wang,
\newblock ``Dual-branch attention-in-attention transformer for single-channel speech enhancement,''
\newblock in {\em Proc. IEEE Int. Conf. Acoust. Speech Signal Process.}, 2022, pp. 7847--7851.

\bibitem{dang2022dpt}
Feng Dang, Hangting Chen, and Pengyuan Zhang,
\newblock ``Dpt-fsnet: Dual-path transformer based full-band and sub-band fusion network for speech enhancement,''
\newblock in {\em Proc. IEEE Int. Conf. Acoust. Speech Signal Process.}, 2022, pp. 6857--6861.

\bibitem{hao2022fast}
Xiang Hao and Xiaofei Li,
\newblock ``Fast fullsubnet: Accelerate full-band and sub-band fusion model for single-channel speech enhancement,''
\newblock {\em 2022, arXiv:2212.09019}.

\bibitem{chen2023ultra}
Hangting Chen, Jianwei Yu, Yi~Luo, Rongzhi Gu, Weihua Li, Zhuocheng Lu, and Chao Weng,
\newblock ``Ultra dual-path compression for joint echo cancellation and noise suppression,''
\newblock {\em 2023, arXiv:2308.11053}.

\bibitem{yu2022high}
Jianwei Yu, Yi~Luo, Hangting Chen, Rongzhi Gu, and Chao Weng,
\newblock ``High fidelity speech enhancement with band-split rnn,''
\newblock {\em 2022, arXiv:2212.00406}.

\bibitem{tu2022maxim}
Zhengzhong Tu, Hossein Talebi, Han Zhang, Feng Yang, Peyman Milanfar, Alan Bovik, and Yinxiao Li,
\newblock ``Maxim: Multi-axis mlp for image processing,''
\newblock in {\em Proc. IEEE Conf. Comput. Vis. Pattern Recognit.}, 2022, pp. 5769--5780.

\bibitem{kuech2014state}
Fabian Kuech, Edwin Mabande, and Gerald Enzner,
\newblock ``State-space architecture of the partitioned-block-based acoustic echo controller,''
\newblock in {\em Proc. IEEE Int. Conf. Acoust. Speech Signal Process.}, 2014, pp. 1295--1299.

\bibitem{liu2021pay}
Hanxiao Liu, Zihang Dai, David So, and Quoc~V Le,
\newblock ``Pay attention to mlps,''
\newblock {\em Proc. Adv. Neural Inf. Process. Syst.}, vol. 34, pp. 9204--9215, 2021.

\bibitem{huang2017densely}
Gao Huang, Zhuang Liu, Laurens Van Der~Maaten, and Kilian~Q Weinberger,
\newblock ``Densely connected convolutional networks,''
\newblock in {\em Proc. IEEE Conf. Comput. Vis. Pattern Recognit.}, 2017, pp. 4700--4708.

\bibitem{schroter2022deepfilternet}
Hendrik Schroter, Alberto~N Escalante-B, Tobias Rosenkranz, and Andreas Maier,
\newblock ``Deepfilternet: A low complexity speech enhancement framework for full-band audio based on deep filtering,''
\newblock in {\em Proc. IEEE Int. Conf. Acoust. Speech Signal Process.}, 2022, pp. 7407--7411.

\bibitem{panayotov2015librispeech}
Vassil Panayotov, Guoguo Chen, Daniel Povey, and Sanjeev Khudanpur,
\newblock ``Librispeech: an asr corpus based on public domain audio books,''
\newblock in {\em Proc. IEEE Int. Conf. Acoust. Speech Signal Process.}, 2015, pp. 5206--5210.

\bibitem{reddy2021interspeech}
Chandan~KA Reddy, Harishchandra Dubey, Kazuhito Koishida, Arun Nair, Vishak Gopal, Ross Cutler, Sebastian Braun, Hannes Gamper, Robert Aichner, and Sriram Srinivasan,
\newblock ``Interspeech 2021 deep noise suppression challenge,''
\newblock {\em 2021, arXiv:2101.01902}.

\bibitem{ko2017study}
Tom Ko, Vijayaditya Peddinti, Daniel Povey, Michael~L Seltzer, and Sanjeev Khudanpur,
\newblock ``A study on data augmentation of reverberant speech for robust speech recognition,''
\newblock in {\em Proc. IEEE Int. Conf. Acoust. Speech Signal Process.}, 2017, pp. 5220--5224.

\bibitem{cutler2022icassp}
Ross Cutler, Ando Saabas, Tanel Parnamaa, Marju Purin, Hannes Gamper, Sebastian Braun, Karsten S{\o}rensen, and Robert Aichner,
\newblock ``Icassp 2022 acoustic echo cancellation challenge,''
\newblock in {\em Proc. IEEE Int. Conf. Acoust. Speech Signal Process.}, 2022, pp. 9107--9111.

\bibitem{luo2019conv}
Yi~Luo and Nima Mesgarani,
\newblock ``Conv-tasnet: Surpassing ideal time--frequency magnitude masking for speech separation,''
\newblock {\em IEEE Trans. Audio Speech Lang. Process.}, vol. 27, no. 8, pp. 1256--1266, 2019.

\bibitem{rec2005p}
ITUT Rec,
\newblock ``P. 862.2: Wideband extension to recommendation p. 862 for the assessment of wideband telephone networks and speech codecs,''
\newblock {\em Int. Telecommu. Uni.}, 2005.

\bibitem{theodoridis2013academic}
Sergios Theodoridis and Rama Chellappa,
\newblock {\em Academic press library in signal processing: Image, video processing and analysis, hardware, audio, acoustic and speech processing},
\newblock Academic Press, 2013.

\bibitem{purin2022aecmos}
Marju Purin, Sten Sootla, Mateja Sponza, Ando Saabas, and Ross Cutler,
\newblock ``Aecmos: A speech quality assessment metric for echo impairment,''
\newblock in {\em Proc. IEEE Int. Conf. Acoust. Speech Signal Process.}, 2022, pp. 901--905.

\end{thebibliography}

\end{document}